\documentclass[11pt,letterpaper]{article}

\usepackage[T1]{fontenc}
\usepackage[utf8]{inputenc}
\usepackage{XCharter}
\usepackage[scaled=0.94]{sourcesanspro}
\usepackage[scaled=0.82]{sourcecodepro}
\usepackage{microtype}

\usepackage[letterpaper,top=1.15in,bottom=1.15in,left=1.35in,right=1.35in,
            headsep=16pt,footskip=30pt]{geometry}
\usepackage{booktabs}
\usepackage{longtable}
\usepackage{array}
\usepackage{calc}
\usepackage{float}
\usepackage{enumitem}
\usepackage{titlesec}
\usepackage{titletoc}
\usepackage{fancyhdr}
\usepackage{etoolbox}
\usepackage[table]{xcolor}
\usepackage[hidelinks]{hyperref}

\definecolor{accent}{HTML}{14507A}   
\definecolor{ink}{HTML}{1A1A1A}      
\definecolor{muted}{HTML}{6B6B6B}    
\definecolor{hairline}{HTML}{C8D4DE} 

\hypersetup{
  colorlinks=true, linkcolor=accent, urlcolor=accent, citecolor=accent,
  pdftitle={AI Can Do Your Homework. Now What?},
  pdfauthor={Akbar, Challen, Fund, Hopkins, Karnalim, Lin, McGuffee, Taneja, Ware},
}

\newlength{\authorblockwidth}

\color{ink}
\providecommand{\tightlist}{\setlength{\itemsep}{0pt}\setlength{\parskip}{0pt}}

\newtoggle{justpart}
\togglefalse{justpart}

\titleformat{\section}[display]
  {\clearpage\normalfont\sffamily\huge\bfseries\color{accent}}
  {}{0pt}{}
  [\vspace{0.9ex}{\color{hairline}\titlerule[1.2pt]}\global\toggletrue{justpart}]
\titlespacing*{\section}{0pt}{0pt}{2.4ex}

\titleformat{\subsection}
  {\iftoggle{justpart}{\global\togglefalse{justpart}}{\clearpage}%
   \normalfont\sffamily\LARGE\bfseries\color{accent}}
  {}{0pt}{}
\titlespacing*{\subsection}{0pt}{0pt}{1.8ex}

\titleformat{\subsubsection}
  {\normalfont\sffamily\Large\bfseries\color{accent}}
  {}{0pt}{}
\titlespacing*{\subsubsection}{0pt}{2.6ex}{0.8ex}

\fancypagestyle{plain}{\fancyhf{}\fancyfoot[C]{\sffamily\footnotesize\color{muted}\thepage}}

\setlist[itemize]{leftmargin=1.3em, itemsep=0.3ex, topsep=0.7ex, parsep=0pt}
\setlist[enumerate]{leftmargin=1.6em, itemsep=0.3ex, topsep=0.7ex, parsep=0pt}

\renewcommand{\arraystretch}{1.25}
\renewcommand{\toprule}{\arrayrulecolor{accent}\specialrule{1pt}{0pt}{0pt}\arrayrulecolor{black}}
\renewcommand{\midrule}{\arrayrulecolor{hairline}\specialrule{0.5pt}{2pt}{2pt}\arrayrulecolor{black}}
\renewcommand{\bottomrule}{\arrayrulecolor{accent}\specialrule{1pt}{2pt}{0pt}\arrayrulecolor{black}}

\newcommand{\pandocrule}{}

\usepackage{tocloft}

\begin{document}

\begin{titlepage}
\thispagestyle{empty}
\centering
\vspace*{-0.55in}
{\color{hairline}\rule{2.2in}{1.2pt}}\par
\vspace{1.1em}
{\sffamily\fontsize{21}{25}\selectfont\bfseries\color{accent}
 AI Can Do Your Homework.\\[0.15em] Now What?\par}
\vspace{0.9em}
{\large\color{muted} Report from an online workshop on computing\\
 assessment in the age of generative AI\par}
\vspace{0.8em}
{\footnotesize\color{muted} \href{https://sigcsevirtual2026.acm.org/track/sigcse-virtual-2026-working-groups}{SIGCSE Virtual Working Group}: \textit{A Taxonomy of Computing Assignments Designed for GenAI Use}\par}
\vspace{0.9em}
{\color{hairline}\rule{2.2in}{1.2pt}}\par
\vspace{1.1em}

{\sffamily\footnotesize\color{accent}\MakeUppercase{Working Group Members}\par}
\vspace{0.6em}
{\footnotesize\renewcommand{\arraystretch}{1.05}
\begin{minipage}{\authorblockwidth}
\begin{tabular}{@{}l@{}}
Muhammad Sajjad Akbar{\color{muted}, \href{https://www.sydney.edu.au/}{\textcolor{muted}{University of Sydney \texttt{(sydney.edu.au)}}}} \\
Geoffrey Challen{\color{muted}, \href{https://illinois.edu/}{\textcolor{muted}{University of Illinois Urbana-Champaign \texttt{(illinois.edu)}}}} \\
Fraida Fund{\color{muted}, \href{https://engineering.nyu.edu/}{\textcolor{muted}{NYU Tandon School of Engineering \texttt{(engineering.nyu.edu)}}}} \\
Casey Hopkins{\color{muted}, \href{https://www.swansea.ac.uk/}{\textcolor{muted}{Swansea University \texttt{(swansea.ac.uk)}}}} \\
Oscar Karnalim{\color{muted}, \href{https://www.maranatha.edu/}{\textcolor{muted}{Maranatha Christian University \texttt{(maranatha.edu)}}}} \\
Kevin Lin{\color{muted}, \href{https://www.washington.edu/}{\textcolor{muted}{University of Washington \texttt{(washington.edu)}}}} \\
James McGuffee{\color{muted}, \href{https://www.ccis.edu/}{\textcolor{muted}{Columbia College (Missouri) \texttt{(ccis.edu)}}}} \\
Shubbhi Taneja{\color{muted}, \href{https://www.wpi.edu/}{\textcolor{muted}{Worcester Polytechnic Institute \texttt{(wpi.edu)}}}} \\
Ranysha Ware{\color{muted}, \href{https://www.swarthmore.edu/}{\textcolor{muted}{Swarthmore College \texttt{(swarthmore.edu)}}}}
\end{tabular}
\end{minipage}\par}

\vspace{1.0em}
\begin{minipage}{\authorblockwidth}
\centering
{\sffamily\footnotesize\color{accent}\MakeUppercase{Workshop Participants}\par}
\vspace{0.6em}
{\footnotesize
\begin{minipage}{\authorblockwidth}\raggedright
\setlength{\parskip}{0.05em}
\hangindent=1.2em\hangafter=1 Barbara Anthony{\color{muted}, \href{https://www.southwestern.edu/}{\textcolor{muted}{Southwestern University \texttt{(southwestern.edu)}}}}\par
\hangindent=1.2em\hangafter=1 Paola Bruscoli{\color{muted}, \href{https://www.bath.ac.uk/}{\textcolor{muted}{University of Bath \texttt{(bath.ac.uk)}}}}\par
\hangindent=1.2em\hangafter=1 David P. Bunde{\color{muted}, \href{https://www.knox.edu/}{\textcolor{muted}{Knox College \texttt{(knox.edu)}}}}\par
\hangindent=1.2em\hangafter=1 Barry Burd{\color{muted}, \href{https://www.drew.edu/}{\textcolor{muted}{Drew University \texttt{(drew.edu)}}}}\par
\hangindent=1.2em\hangafter=1 Michaël Cadilhac{\color{muted}, \href{https://www.depaul.edu/}{\textcolor{muted}{DePaul University \texttt{(depaul.edu)}}}}\par
\hangindent=1.2em\hangafter=1 Dan DeBlasio{\color{muted}, \href{https://www.cmu.edu/}{\textcolor{muted}{Carnegie Mellon University \texttt{(cmu.edu)}}}}\par
\hangindent=1.2em\hangafter=1 Kutub Gandhi{\color{muted}, \href{https://www.ut.edu/}{\textcolor{muted}{University of Tampa \texttt{(ut.edu)}}}}\par
\hangindent=1.2em\hangafter=1 Nadeem Abdul Hamid{\color{muted}, \href{https://www.berry.edu/}{\textcolor{muted}{Berry College \texttt{(berry.edu)}}}}\par
\hangindent=1.2em\hangafter=1 Brian K. Hare{\color{muted}, \href{https://www.umkc.edu/}{\textcolor{muted}{University of Missouri-Kansas City \texttt{(umkc.edu)}}}}\par
\hangindent=1.2em\hangafter=1 R. Bruce Irvin{\color{muted}, \href{https://www.pdx.edu/}{\textcolor{muted}{Portland State University \texttt{(pdx.edu)}}}}\par
\hangindent=1.2em\hangafter=1 Ying Li{\color{muted}, \href{https://www.colby.edu/}{\textcolor{muted}{Colby College \texttt{(colby.edu)}}}}\par
\hangindent=1.2em\hangafter=1 Dave Musicant{\color{muted}, \href{https://www.carleton.edu/}{\textcolor{muted}{Carleton College \texttt{(carleton.edu)}}}}\par
\hangindent=1.2em\hangafter=1 S. Monisha Pulimood{\color{muted}, \href{https://www.tcnj.edu/}{\textcolor{muted}{The College of New Jersey \texttt{(tcnj.edu)}}}}\par
\hangindent=1.2em\hangafter=1 Karen Reid{\color{muted}, \href{https://www.utoronto.ca/}{\textcolor{muted}{University of Toronto \texttt{(utoronto.ca)}}}}\par
\hangindent=1.2em\hangafter=1 Mahsa Sadeghi Ghahroudi{\color{muted}, \href{https://www.dawsoncollege.qc.ca/}{\textcolor{muted}{Dawson College \texttt{(dawsoncollege.qc.ca)}}}}\par
\hangindent=1.2em\hangafter=1 Md Kamruzzaman Sarker{\color{muted}, \href{https://www.bowiestate.edu/}{\textcolor{muted}{Bowie State University \texttt{(bowiestate.edu)}}}}\par
\hangindent=1.2em\hangafter=1 James Vanderhyde{\color{muted}, \href{https://www.sxu.edu/}{\textcolor{muted}{Saint Xavier University \texttt{(sxu.edu)}}}}\par
\hangindent=1.2em\hangafter=1 Debbie S. Yuster{\color{muted}, \href{https://www.ramapo.edu/}{\textcolor{muted}{Ramapo College of New Jersey \texttt{(ramapo.edu)}}}}\par
\end{minipage}\par}
\vspace{1.1em}
{\footnotesize\color{muted}
\begin{minipage}{\authorblockwidth}\raggedright
Participants are listed here at their own request.
Listing was optional and opt-in, and no idea in this report is attributed to any named individual.
\end{minipage}\par}
\end{minipage}\par

\vspace{1.1em}
{\sffamily\footnotesize\color{muted}Workshop held online, 28 July 2026 \textperiodcentered{} Report revised 18 August 2026\par}
\vfill
\end{titlepage}

\tableofcontents
\clearpage

\subsection{Introduction}\label{introduction}

This report summarizes a two-hour online workshop held on 28 July 2026,
convened by the SIGCSE Virtual Working Group studying how computing
educators are adapting assignments in response to generative AI. It is
not a research paper. It is an account of what seventy or so computing
educators said to each other when given two hours and no agenda beyond a
set of questions they had written themselves.

\subsubsection{Why the workshop was designed this
way}\label{why-the-workshop-was-designed-this-way}

The working group's premise is that assignments are where the problem of
generative AI in computing education is actually being felt. A
department can adopt a policy in a meeting; an instructor has to decide
what to put in front of students in the fall. We wanted contact with
people making that decision, and we wanted it while they were still
making it.

Three design choices followed. First, discussion rather than talks: no
presentations, no slides, and nothing for attendees to prepare. Second,
rooms organized around concrete adaptation strategies, on the reasoning
that educators already experience the problem in those terms---whether
to move to open-ended tasks, whether to bring back proctored exams---so
strategies make the easiest on-ramp to a real conversation. Third, two
rounds: an hour organized by strategy, then an hour organized by
cross-cutting questions, so that participants met people who had made
different choices from their own.

The cost of this design is that it produces no data in the usual sense.
Nothing here is measured, sampled, or controlled. What it produces
instead is a record of practice at a moment when published literature is
twelve to eighteen months behind the tools, and of the specific ways
experienced instructors are finding their existing methods failing.

\subsubsection{Who came, and what they wanted to
discuss}\label{who-came-and-what-they-wanted-to-discuss}

Registration was open through SIGCSE channels and drew \textbf{118
registrants} by the time the workshop ran. \textbf{Seventy-three people
joined} the session, of whom 64 were participants and nine were working
group members; peak concurrent attendance was around 70, and the median
participant stayed the full two hours. Registration asked which single
strategy each person most wanted to discuss and, in an open text field,
what question they brought. All but one registrant submitted a question;
forty-seven also submitted questions about strategies other than their
own.

The distribution below reflects what registrants said they wanted to
discuss, not who turned up. Room structure was set a day earlier, from
the snapshot available on 27 July, and the ordering did not change
materially in the final day of sign-ups:

{\def\LTcaptype{none} 
\begin{table}[H]
\centering
\begin{tabular}{@{}lr@{}}
\toprule
Strategy & Registrants \\
\midrule
Oral \& Interactive Assessment & 21 \\
Open-Ended \& Authentic Task Design & 20 \\
Rubric \& Grading Redesign & 19 \\
Ambitious, AI-Leveraged Projects & 17 \\
Evaluating \& Fixing AI-Produced Work & 15 \\
Controlled \& AI-Free Assessment & 13 \\
Process Evidence \& Effort Signals & 12 \\
Transparent AI Use \& Literacy & 1 \\
\bottomrule
\end{tabular}
\end{table}
}

Seven strategies drew enough interest to run as rooms. Transparent AI
Use and Literacy drew a single registrant and was not run separately;
that person joined one of the other rooms. Its concerns surfaced anyway,
mostly in the rubrics and process evidence rooms, and disclosure
practices appear throughout this report.

Two things about that distribution are worth noting. The first is how
flat it is: the seven strategies that ran span 21 registrants down to
12, with no strategy drawing even a fifth of the total. That tells us
interest was spread across every strategy on offer rather than
concentrated on one. It does not tell us whether the field agrees about
what to do---a group of people holding firm but opposing views would
produce the same flat spread. The evidence that the field has not agreed
is in Part III, where the same unresolved questions surface room after
room, whichever strategy the room was organized around.

The second is that interest divides almost evenly between two quite
different responses. Redesigning what gets assigned---open-ended and
authentic tasks, ambitious AI-leveraged projects---drew 37 registrants.
Redesigning how the work is verified or valued---oral assessment,
rubrics and grading---drew 40. Instructors are pursuing both at once
rather than choosing between them, which is consistent with what the
rooms themselves sounded like.

The clear outlier is Transparent AI Use and Literacy, the only category
about teaching AI itself rather than about assessing work produced with
it. In a workshop convened around generative AI, it drew one registrant.

\subsubsection{How the two hours ran}\label{how-the-two-hours-ran}

Roughly five minutes of welcome, fifty minutes in strategy rooms, a
ten-minute break, fifty minutes in question rooms, and a short close.
Participants self-selected into rooms and could move between them,
though we did not capture how often anyone did; the room accounts below
reflect whoever was present when something was said.

Each room had a working group member as facilitator, and each opened
with the questions its own registrants had submitted.

Eight question rooms were planned for the second hour and seven ran; the
room on trustworthy process evidence was not opened, and its facilitator
joined the rubrics room instead.

\subsubsection{How this report was made}\label{how-this-report-was-made}

Each room had a tab in a shared document, open to all participants, in
which the facilitator took notes and anyone could add. The session was
not recorded. Zoom's live transcript ran only to support note-taking,
and some facilitators kept fuller transcripts of their own rooms.

The sections that follow were written from those notes and transcripts.
Where a facilitator wrote a narrative summary of their own room, that
text forms the body of the section, edited lightly and augmented from
the room's raw notes and transcript where those contained specifics the
summary had condensed away. Where a facilitator left detailed notes
instead, the section was written from those. Sections were deliberately
evened out in length so that a room reads according to what was said in
it rather than according to how much its facilitator happened to write
down. Material that arrived independently in several rooms was moved to
Part III rather than repeated, and cross-references to other rooms were
added by the editors, since no facilitator could see beyond their own.

The record is anonymous by default. No idea in this report is attributed
to a named individual, and institutional details have been removed or
generalized where they would identify a participant or a student.
Participants may opt to be listed as contributors to the report as a
whole; that list is collected after this draft circulates.

\subsubsection{How to read this report}\label{how-to-read-this-report}

Part I contains one section per adaptation strategy, following the seven
Hour 1 breakout rooms. Each section absorbs the Hour 2 question room
that was seeded from it, since those pairings were deliberate and
several facilitators summarized both of their rooms together. Part II
covers the two Hour 2 rooms that were cross-cutting by design and had no
strategy pair. Part III steps back: it collects the arguments that
arrived independently in several rooms, and the patterns that only
become visible across the whole workshop.

Readers wanting concrete practice will find most of it in Part I.
Readers wanting the argument the field is currently having with itself
will find it in Part III.

Every section opens with the strategy as we defined it for the workshop,
lists the questions that were seeded for discussion, and then summarizes
what the room actually discussed. The seeded questions came from
registrants themselves, lightly edited and de-duplicated; they are
reproduced here because they are themselves a finding about what the
community is worried about.

\pandocrule

\section{Part I: Strategies}\label{part-i-strategies}

\subsection{1. Open-Ended \& Authentic Task
Design}\label{open-ended-authentic-task-design}

\textbf{Facilitator:} Geoffrey Challen, University of Illinois
Urbana-Champaign

\emph{This section covers both the Hour 1 strategy room and the Hour 2
question room ``How do we design authentic, open-ended tasks that hold
up?''}

\subsubsection{The strategy}\label{the-strategy}

Assignments that specify what a student must demonstrate rather than
what they must build. Instead of a problem statement a model can consume
whole, the task names a target---a communication protocol, a form of
concurrency, a comparison worth making---and leaves the student to
decide what to build and how. The facilitator offered an analogy from
photography school, where the assignment is never ``take a picture of
X'' but rather ``motion,'' and the student produces images demonstrating
they can capture it.

\subsubsection{Seeded questions}\label{seeded-questions}

\begin{enumerate}
\def\labelenumi{\arabic{enumi}.}
\tightlist
\item
  How do we lay the knowledge groundwork so students still learn the
  fundamentals through an open-ended, AI-assisted project?
\item
  How do we make a task specific enough that AI can't just do the whole
  thing, while keeping it open-ended?
\item
  How do we scaffold real-world projects to stay accessible across
  student levels, and pick projects that fit different course levels?
\item
  How do we balance open-ended vs.~specified parts to get the right
  amount of student effort?
\item
  How do we do this at scale and still assess technical quality, not
  just the presentation?
\item
  When AI can do the hardest tasks we can set, is prompting all that's
  left, and how do we evaluate the human contribution?
\item
  What makes an open-ended task resist ``just have AI do it''?
\end{enumerate}

\subsubsection{Discussion}\label{discussion}

Two worked examples anchored the first hour, both already run with real
students.

The largest was an open-ended final project in a 700-student systems
programming course. Rather than grade code, the instructor set technical
specifications---a multi-process design using either pipes or sockets
with \texttt{select}---and graded a written document in which students
explained how they met them, identified their communication protocol and
their source of concurrency, and described how they tested, accompanied
by a demo video under five minutes. A script compiled every submission
on the department server to catch projects that did not build at all.
Roughly half the class produced work slightly easier than a typical
final assignment and half met or exceeded expectations. The costs were
substantial: an enormous amount of instructor grading time, most of it
spent working out how to coach TAs through an unfamiliar rubric, and
prolonged negotiation with students who all wanted As. Two observations
landed hardest. Failure rates rose on the course's in-person tests, and
the instructor's reading was not that students were cheating but that
they ``deluded themselves into thinking they were learning when they
weren't.'' And on finally inspecting the code, it was poor---clearly not
meeting the specification---in ways she could spot immediately but
doubted her TAs could without significant training. Videos, by contrast,
were mostly genuine; many scripts were AI-written, but students had put
enough thought in for them to make sense.

The second example inverted the usual worry about scope. An advanced
programming course deliberately assigned more than students could
finish, on the explicit understanding that they would use AI. The
project was a real commercial system, built originally by about a
hundred programmers over roughly a year, scaled down so students
received about half, divided into three segments with both directions of
each: API to database, database to API, API to screen. The design
exploited a limitation the instructor had observed---that models handled
two coupled concerns poorly---so a single well-formed request would not
produce a solution and students had to decompose before delegating. The
project ran the whole semester from day one, replacing the per-language
assignments the course previously used, a change cleared with the
department chair in advance. Ten minutes of each class became a
recurring negotiation over what could be cut, with the instructor
holding the line that whatever remained had to hang together as a
working system---putting students inside the scoping decision rather
than receiving it. Two teams kept working into the following semester
for no credit. Every assignment carries an AI disclosure at the top, and
any AI use must be cited like a book or article. Asked how students
engaged, the instructor noted the familiar asymmetry---the strongest
students never come to office hours---then arrived at a conclusion he
said he had not articulated before that moment: what he had actually
been teaching was problem decomposition.

Both rooms spent substantial time on a question larger than task
design---whether students still need the fundamentals, and what those
now are. Because the same argument surfaced independently in four rooms,
it is treated in the cross-cutting chapter rather than here.

The second hour opened by questioning whether the question was well
posed. If we do not know where model capability will land, one
participant argued, the durable test is not whether a task resists AI
but whether there is a benefit to asking it at all---a criterion that
survives whatever the tools turn out to do. Another was blunter:
AI-resistant programming assignments are no longer possible. The
facilitator's amendment was that they are possible but temporary,
lasting perhaps a semester. That reframing mattered for the rest of the
discussion. If resistance has a shelf life, the design goal is not a
task AI cannot do but a task whose difficulty survives AI doing part of
it.

Underneath the second hour ran a structural question: whether any of
this can be answered at the level of a single course. The facilitator
argued that it demands a curricular response rather than a course
response. An instructor can optimize their own piece, but a course that
changes while those downstream do not leaves students crossing an
inconsistent boundary and the instructor who changed absorbing the
friction alone. Participants described exactly that inconsistency on
their own faculties---some colleagues permitting AI freely as simply
another tool, others requiring students to separate their own work from
the model's, and no departmental position reconciling them. The obstacle
identified was speed: curricular conversations have begun at several
institutions, but nobody expected them to conclude quickly enough to
help the students currently moving through their programs, which leaves
individual instructors making local decisions whose costs are also
local.

The concrete design ideas were smaller than the theorizing but more
usable. An open-ended CS3 text-processing assignment gives students an
unstructured corpus---the CIA World Factbook, or a historical
dataset---and asks them to find the structure, choose which entities and
categories to compare, and decide what story to tell, writing their own
tests along the way. Because the requirements are the student's to
invent, AI does not remove the hard part; another participant reframed
it as putting the student in the customer role with AI as the
programmer. A second idea was a comparative design assignment: have a
model implement the same project in several design patterns, then ask
students to analyze the differences and draw the design diagrams. Both
share the premise the room kept returning to---when the specification is
the student's responsibility, the work AI cannot absorb is the work of
deciding what to build.

\pandocrule

\subsection{2. Ambitious, AI-Leveraged
Projects}\label{ambitious-ai-leveraged-projects}

\textbf{Facilitator:} Fraida Fund, NYU Tandon School of Engineering

\subsubsection{The strategy}\label{the-strategy-1}

Assignments that deliberately exceed what a student could complete
unaided, and that treat AI collaboration as a premise of the work rather
than a concession to it. Students collaborate with AI agents on tasks
scoped beyond their unassisted reach, on the premise that if AI raises
the floor, the ceiling should rise with it: rather than shrinking
assignments until AI can no longer complete them, make them larger than
either the student or the model could manage alone.

\subsubsection{Seeded questions}\label{seeded-questions-1}

\begin{enumerate}
\def\labelenumi{\arabic{enumi}.}
\tightlist
\item
  How should an AI-leveraged project be structured?
\item
  How do we manage access and control tool costs for students and
  institutions?
\item
  How do we design a project that requires students to use AI to learn,
  but that AI can't fully solve for them?
\item
  How do we grade an ambitious project, and assess the higher-order work
  when AI handles the foundations?
\item
  If a project is scaffolded, what stops a student from just handing the
  whole thing back to AI?
\item
  How do we build semester-long projects students can actually showcase
  (portfolio and industry alignment)?
\item
  What's the role of LLM ``skills'' and ``agents,'' and how do we teach
  responsible, reflective AI use?
\end{enumerate}

\subsubsection{Discussion}\label{discussion-1}

The facilitator noted before the session that the questions submitted by
this room's registrants highlighted two tensions in particular: using
powerful tools versus managing costs, and underreliance versus
appropriate reliance versus overreliance. Both ran through the
discussion that followed.

Participants are guiding students in using AI assistance to undertake
ambitious projects, such as working with large codebases, implementing
software from specifications, modifying existing open-source software,
and reproducing published research. A networking course had students use
AI to set up experiments so they could engage with current research
without full mastery of the tooling; students felt they contributed far
more than they could have unassisted, though how much they learned, and
how to quantify it, remained open. A machine learning systems course
asked students to implement an ML feature inside an existing open-source
project, on the reasoning that the instructor never expected familiarity
with that codebase or its languages. A requirements engineering course
had students derive specifications and then implement them with coding
agents.

These projects typically involve tasks that students would not be
expected to complete without AI assistance. Because AI use is integral
to the work, this raises questions about ensuring equitable access and
managing costs. Participants discussed several strategies, including
using ``bring your own provider'' agent harnesses rather than tools tied
to specific subscription plans, arranging access to institutionally
supported models or showing students how to access high-quality free
models, and using research infrastructure such as CloudLab, FABRIC, and
Chameleon. An advantage noted for the harness approach is that students
can keep using it with free models after the course ends.

This type of use also raises questions about how to accommodate students
with ethical, sustainability, or privacy concerns related to generative
AI---concerns raised in the room included energy and water use,
community impact, the provenance of training data, and privacy. These
students may be directed toward more acceptable alternatives, such as
small, self-hosted models trained on open data. Because these models are
often less capable, however, students may directly experience the
tradeoffs between ethical considerations and model performance. This
makes it especially important to evaluate students on how effectively
they frame problems, construct prompts, and critically guide the models,
rather than primarily on the quality of the models' outputs.

Evaluation emerged as a central challenge when students are expected to
use generative AI. Participants described combining assessment of final
deliverables with presentations, oral examinations, demonstrations,
peer-review-style feedback, and qualitative reports. Several emphasized
asking students to explain how their code works, justify key decisions,
identify alternatives, and respond to counterfactual questions in order
to assess whether they had meaningfully engaged with the project rather
than simply produced a strong artifact. One vertically integrated
project program runs weekly presentations to sub-teams of TAs and
mentors, correcting course throughout the semester rather than at the
end.

At the same time, these evaluation approaches introduce practical and
conceptual difficulties. Specifications-based grading may also
unintentionally encourage students to submit the rubric directly to an
AI system, without meaningful engagement or learning. Oral and
interview-based evaluation can be difficult to scale, and a student's
inability to explain a system may reflect weak understanding, weak
communication skills, or both---a distinction some participants observed
has become harder, with students breaking down on counterfactual
questions even when they had done the work. Client- or sponsor-facing
projects provide an additional source of accountability and may mitigate
scalability concerns to some extent, but external satisfaction does not
necessarily demonstrate that students understand the work. These
concerns suggest that evaluation should examine not only the quality of
the final output, but also students' reasoning, decision-making, ability
to critique results, and capacity to explain and adapt their work.

The discussion also highlighted concerns about both overreliance on and
underuse of AI tools. Overreliance may prevent students from developing
core problem-solving skills, particularly if they have little experience
working through complex tasks without AI assistance. At the same time,
students may underuse AI by defaulting to familiar, low-friction chat
interfaces or struggling to collaborate effectively with coding agents.
Using AI effectively is not simply a way to avoid work; it requires
substantial effort to articulate goals, delegate tasks, evaluate
results, provide feedback, and integrate AI-generated contributions.

Participants suggested that effective AI use may depend on skills that
have not traditionally been central to computing education, including
reading and evaluating unfamiliar code, articulating objectives,
decomposing work, providing useful feedback, and managing an agent's
progress. In this sense, working effectively with AI may resemble
managing a collaborator or research team. The group did not identify a
clear solution, but noted that the skills associated with success in an
AI-supported environment may differ from those previously emphasized in
computer science education, making it difficult to determine the
appropriate balance between independent work and AI-assisted
collaboration. This challenge is compounded by a fundamental asymmetry
between educators and students: many educators developed their technical
foundations before AI tools were widely available and only later learned
to work with them, while today's students cannot replicate that
sequence. Educators therefore have limited firsthand experience of what
it means to build foundational skills while AI is present from the
outset, making it harder to judge which forms of struggle remain
necessary and which can productively be delegated to AI.

\pandocrule

\subsection{3. Evaluating \& Fixing AI-Produced
Work}\label{evaluating-fixing-ai-produced-work}

\textbf{Facilitator:} Casey Hopkins, Swansea University

\emph{This section covers both the Hour 1 strategy room and the Hour 2
question room ``How do we teach and grade the critique of AI output?'',
which the facilitator summarized together.}

\subsubsection{The strategy}\label{the-strategy-2}

Assignments in which the object of study is output the student did not
write. Students read, test, critique, debug, and repair AI-produced code
rather than producing it from scratch. The wager is that judgment about
code survives as an assessable skill even when the production of code
does not.

\subsubsection{Seeded questions}\label{seeded-questions-2}

\begin{enumerate}
\def\labelenumi{\arabic{enumi}.}
\tightlist
\item
  How do we do critique-of-AI-output at scale, e.g., 400+ students and a
  handful of TAs?
\item
  How do we design the assignment so the student does the critique, not
  the AI?
\item
  AI does so well on introductory topics that there's little to
  critique---how do we handle that?
\item
  When should critique enter, and does it differ for code generation
  vs.~design/architecture?
\item
  Vibe-coding produces a lot of extraneous code; should students be able
  to explain or strip it away?
\item
  What skills matter most to teach once AI can do the basic coding?
\end{enumerate}

\subsubsection{Discussion}\label{discussion-2}

The discussion explored how computing educators can teach and assess
students' ability to evaluate, critique, and improve AI-produced work.
Rather than focusing on limiting AI use, participants discussed how to
design learning activities and assessments that develop students'
judgement, critical thinking, and software engineering skills when
working with AI-generated outputs. Participants reported that students
are already using these tools throughout the curriculum regardless of
course policy, in extreme cases without reading the output at all.

A key theme was trying to ensure that students, rather than AI, perform
the evaluation. Participants agreed that preventing AI use is neither
realistic nor desirable; instead, education should focus on helping
students understand when AI performs well, where it fails, and how to
assess its outputs critically. The room was candid that we are past the
point of reliably telling who did what: when one participant asked for
transcripts of students' evaluations, students had the AI produce the
evaluations too. Whatever text you hand a student can become a prompt.
The practical responses were therefore structural rather than
forensic---moving critique into settings where the tool is absent or
irrelevant: in-class and proctored work, debates, oral presentations,
whiteboard exams.

Participants reflected that students frequently question why they should
develop skills that AI appears capable of performing, for example
coding. Educators therefore need to explain that professional software
engineers are expected to evaluate, debug, test, and justify solutions,
regardless of whether those solutions originate from AI or humans---and
this is difficult to do if you don't understand the fundamentals of
coding itself. Messaging emerged as a theme in its own right: if
assignments are split into AI-permitted and AI-free parts, the reason
for the split has to be communicated, or students remain cynical about
why they are doing work the machine could do. Participants observed that
industry is in the same state of uncertainty, which makes the appeal to
professional norms harder to make cleanly, though job interviews remain
a concrete and motivating anchor.

A significant discussion centred on whether students must first learn to
program before they can effectively critique code. Some participants
argued that reading code may become a more important skill than writing
it, particularly in an AI-assisted development environment. Others
maintained that meaningful critique still requires a solid understanding
of programming fundamentals and computer science concepts---the analogy
offered was music criticism, which presupposes theory. One suggested
resolution was pedagogical rather than philosophical: give students code
to critique before they have the skills to do it, let the task defeat
them, and use that failure to motivate learning the fundamentals.

When critiquing AI output, participants recognised that AI often
produces good-quality solutions for introductory programming tasks,
making critique more challenging. One suggestion was to deliberately
prompt AI to generate flawed or poor-quality code that students could
then analyse and improve. Unless in a controlled environment however,
these activities are still open to students using AI to complete the
evaluations.

Considering this, participants generally supported combining closed-book
examinations with open, AI-supported assessments. This balance allows
foundational knowledge to be assessed while recognising that AI tools
are becoming part of professional software development.

The group recognised that scaling assessment remains difficult. Oral
examinations provide strong evidence of individual understanding but are
impractical for large cohorts. Group work, class tests, peer review and
debates were discussed as more scalable alternatives, with peer
assessment contributing only a partial percentage of the overall grade
to limit the consequences of inconsistency. For very large cohorts, one
approach discussed was analysing outputs and AI-generated unit tests
rather than reading source directly. Participants also raised broader
considerations, including ensuring that redesigned assessments remain
accessible for students with support needs and acknowledging the tension
between reducing traditional examinations and reintroducing more
in-person assessment to verify understanding.

There was broad agreement that reading, evaluating and critiquing code
should become a core skill introduced earlier in the curriculum and
developed progressively---critique of correctness and style first,
efficiency and design later. Participants suggested introducing software
engineering practices, particularly testing and code review, much
earlier than is common in many curricula. Participants generally viewed
critique as equally applicable to code generation and higher-level
software design, with the nature of the critique becoming more
sophisticated as students progress.

This room's curricular argument should be read alongside Part III, where
a separate room reached a conclusion in tension with it: that reading
code may not be a durable skill either, if code review is increasingly
conducted between models. The case for teaching critique early rests on
an assumption that room questioned.

\subsubsection{Approaches described in the
room}\label{approaches-described-in-the-room}

\begin{enumerate}
\def\labelenumi{\arabic{enumi}.}
\tightlist
\item
  Scaffold AI use across the curriculum, beginning with a supplied
  prompt and gradually shifting responsibility for formulating prompts
  and evaluating results onto the student.
\item
  Give students AI-generated solutions to critique, improve, or test:
  validating behavior, finding defects or unnecessary complexity,
  identifying missing edge cases, designing unit tests, arguing why one
  implementation beats another.
\item
  Pair AI-supported coursework with in-class quizzes or examinations
  that verify individual understanding.
\item
  Use peer review of AI co-produced work, mirroring industry code
  review, as a route to scale.
\item
  Assess the student's analysis of generated code, tests, and software
  quality rather than the code itself.
\item
  Open with a critique task students find hard, using the difficulty to
  motivate the underlying programming skills.
\item
  Assess the product of AI-assisted co-development rather than the
  codebase.
\item
  Retain a final proctored assessment of fundamentals.
\end{enumerate}

\pandocrule

\subsection{4. Controlled \& AI-Free
Assessment}\label{controlled-ai-free-assessment}

\textbf{Facilitator:} Oscar Karnalim, Maranatha Christian University

\emph{This section covers both the Hour 1 strategy room and the Hour 2
question room ``Will students still learn the fundamentals?''}

\subsubsection{The strategy}\label{the-strategy-3}

Assessment conducted under conditions that exclude AI use, whether by
proctoring, by moving work into the classroom, or by designing tasks a
model cannot complete. This is the one strategy in the set that responds
to AI by preserving rather than adapting, and the room treated it less
as a refusal to change than as a question of where in a curriculum
unassisted work still has to be demonstrated.

\subsubsection{Seeded questions}\label{seeded-questions-3}

\begin{enumerate}
\def\labelenumi{\arabic{enumi}.}
\tightlist
\item
  Do proctored, AI-free assessments need to change at all, and if so,
  how?
\item
  What kinds of questions would AI fail to answer, and thus still probe
  real understanding?
\item
  How do we make AI-free assessment painless to grade and fair to weaker
  test-takers?
\item
  Can a scaled assignment for 200 students still be AI-free and test
  real learning?
\item
  How should AI-free expectations differ between a freshman and a senior
  course, and how do we motivate students to learn the basics first?
\item
  How do we keep homework doing its job, building base skills, in the AI
  era?
\item
  What counts as ``the fundamentals'' now, and which must be learned
  without AI?
\item
  How do we design assessments that verify a student can do it
  themselves?
\item
  Where does AI genuinely help learning vs.~short-circuit it?
\end{enumerate}

\subsubsection{Discussion}\label{discussion-3}

Everything reported below comes from participants' own practice rather
than from speculation, with two exceptions noted where they arise. Most
participants in these rooms taught small classes, which shaped what they
found workable; the facilitator did not collect enrollment figures or
outcome data, so the claims here are about what instructors are doing
and what they believe works, not about measured effect.

The room's answer to whether proctored assessment needs to change was a
qualified no: what changes is not the format but how much weight it
carries and where it sits. Participants reported running more of these
assessments than before, driven by concern about AI misuse and its
effect on what one called \emph{cognitive debt}. The operating rule
several described is conditional: where a learning outcome is
compromised by AI use, move that outcome to a proctored AI-free
assessment. Participants had done this in both small and large classes.
It matters most early in a course, when students are building the
foundation everything later depends on, and it is also used simply to
verify that a student holds knowledge, because critical thinking has
become hard to evaluate from submitted work.

On what a model would still fail, participants offered four approaches
they had already deployed. Larger projects and higher-order thinking
assessments, used in software projects and research methods courses.
In-person interviews conducted after submission. Assessments that move
out of the text modality---two participants now require a diagram or a
video in place of written text. And assessments targeting innovation,
originality, and novelty, used in research methods and startup courses,
where the absence of precedent is the point. Two caveats followed.
Models keep improving, so any question set built on current limitations
requires continuous revision. And some questions defeat a student not
because the model cannot answer them but because the student cannot
construct the prompt---which makes ``hard for AI'' a property of the
interaction rather than of the question.

Practicality drew as much attention as validity. Closed-book written
exams on paper, possibly with whiteboard-style oral interviews, were the
baseline, and participants were direct that these remain painful to
grade. One participant had moved to proctoring with digital entry,
having students bring laptops so the work is typed and machine-gradable
while the environment stays controlled.

Asked whether a 200-student assignment can be AI-free and still test
real learning, the room's answer was yes, but tricky. Ensuring
submissions are AI-free is difficult without proctoring or a controlled
environment. AI detectors were ruled out on the basis of classroom
experience rather than principle: participants who had used them
reported many false positives, and AI-generated work is often
indistinguishable from human work. What remains is time-consuming, and
the mitigations participants had actually used were dividing
verification across tutors, applying automated marking where possible,
and simply limiting how often the expensive methods are used.

The room drew a clear line by course level, though not the simple one.
Freshmen should learn the fundamentals, so AI is not encouraged early.
For senior courses the answer depends on the course rather than the
year: a foundational senior-level course may still warrant AI-free
assessment, while an advanced or applied one is better served by
teaching students to leverage AI in that context.

Homework generated the room's most concrete thread, because it is where
the strategy is hardest to sustain, and here participants named the
courses. Setting assessments large enough that AI cannot complete them
alone, done in advanced courses. Using a flipped classroom so lecture
time can be spent working on assignments under supervision. Requiring
students to begin without AI, done in evolutionary computing and machine
learning courses. And micro-sized milestones combined with version
control history to surface the large sudden changes characteristic of
pasted generated code---used in software projects and evolutionary
computing courses, with one participant applying version control this
way on their assignments directly.

Those techniques also answer the two problems the room had otherwise
left open. Large AI-free assignments that will not fit into class time
are addressed by the flipped classroom, which converts lecture time into
supervised working time. Distance students, for whom no controlled
environment exists, are addressed by micro-milestones and version
control---though here participants were offering suggestions rather than
reporting experience, and the problem should be read as open.

Turning to the fundamentals themselves, the room's working definition
was broader than programming ability: foundational concepts, which
participants noted exist at every level from freshman to graduate rather
than only at the start; computational thinking and problem solving; AI
literacy and prompting skill; competence with platforms and supporting
environments; and code and program comprehension. Two shifts were
proposed---less emphasis on building from scratch and more on verifying
and adapting AI output, and the suggestion that memorization may no
longer belong at all, since students can retrieve information from AI
directly.

For verifying that a student can work unaided, participants listed
controlled AI-free environments for small assessments,
micro-scaffolding, interviews, presentations, oral examinations, and
self-reflections used as a preliminary authenticity check rather than as
a graded artifact. Two obstacles were named with their remedies.
Qualitative grading is hard to do consistently, addressed by simplifying
the marking scale or writing detailed rubrics. And oral and presentation
formats disadvantage students with communication limitations, an equity
concern that recurred in several other rooms and that nobody resolved.

Finally, on where AI helps learning rather than short-circuiting it, the
room converged on a claim with an uncomfortable implication. AI is most
useful to students who already understand the material, who have solid
computational thinking, or who understand the characteristics of the
tool well enough to direct it. Generative AI is very good at producing
personalized tutorials, but weaker students need training before they
can benefit from them. The benefit, in other words, accrues fastest to
the students who need it least.

One registrant's question was submitted to this room but never posed,
because time ran out: how do I embrace the reality of vibe coding and at
the same time compel students to learn the fundamentals? The
facilitator's own answer, drawn from the discussion rather than voiced
in it, is the position the room's practice implies. Students need
fundamentals and code comprehension \emph{before} vibe coding, because
they have to be able to verify what the model produces. Those are taught
in introductory courses. Therefore introductory assessments should be
AI-free, or at minimum controlled---which up to three participants in
the room, including the facilitator, already do.

\pandocrule

\subsection{5. Process Evidence \& Effort
Signals}\label{process-evidence-effort-signals}

\textbf{Facilitator:} Kevin Lin, University of Washington \emph{The Hour
2 question room paired with this strategy, ``How do we get process
evidence we can trust?'', was not opened; its facilitator joined the
Rubric \& Grading Redesign room to balance participant numbers. The
questions seeded for it---what proofs of work survive when the process
itself can be AI-generated, how to capture process without creating
busywork or surveillance, and what tools or exports actually help---went
undiscussed as a set, though each surfaced in other rooms.}

\subsubsection{The strategy}\label{the-strategy-4}

Grading how the work was done rather than only what was produced. The
evidence discussed ranged widely: AI interaction logs and transcripts,
drafts and commit history, prompt and iteration portfolios, time on
task, and reflective commentary on AI use. If the artifact is no longer
evidence that learning occurred, the process that produced it becomes
the candidate replacement.

\subsubsection{Seeded questions}\label{seeded-questions-4}

\begin{enumerate}
\def\labelenumi{\arabic{enumi}.}
\tightlist
\item
  If answers are no longer proof of work, what are the better proofs of
  work and learning?
\item
  How do we know process documentation, or a reflection, reflects the
  real process and isn't itself AI-generated?
\item
  How do we shift students from focusing on the artifact to valuing the
  process?
\item
  How do we actually measure and grade process evidence without it
  becoming busywork?
\item
  How do we help students write meaningful prompts, and how much prompt
  engineering are we responsible for teaching?
\item
  How do we make reflection genuine rather than a superficial box-check?
\end{enumerate}

\subsubsection{Discussion}\label{discussion-4}

The room framed the situation as an opportunity before framing it as a
problem. If AI absorbs some of the easier skills and knowledge, courses
can move toward more realistic and professional work, including
questions of agency---asking students to consider what they are
responsible for. Design and requirements work becomes correspondingly
more important. But the balance is genuinely hard to strike:
participants kept returning to the difficulty of determining what
novices need to know in order to do advanced work at all, how to help
students build the independent skills to get there, and how to convey
why a skill matters when a model can perform it. Aligning any of this
with stated learning objectives was described as tricky in its own
right.

Underneath sat the room's central claim: the old approach of accepting
an answer as proof of work no longer holds, which raises the question of
how to teach in an age of answer machines. Participants connected this
to socialization rather than detection---how to bring students to value
the parts of an education that are not about obtaining an answer.
Students' own mindsets were named as an obstacle, with grade-focused
goals interfering with any interest in the process that produces them.
The pace of change compounded it: two years ago models were not good at
producing even small programs, and they are now competent across a wide
range of programming tasks, which makes it hard to adapt within a single
term.

On assessing process directly, participants agreed that oral formats
help but ran immediately into the time cost of doing it with everyone,
and asked what kinds of tasks would actually reveal process. A sharper
objection followed. Some students perform well in an oral format by
studying AI-generated work carefully, without possessing the underlying
skill to have produced it. Short of invigilating a two-hour session, the
room could not identify a reliable way to measure the capacity to do
something without AI.

The most concrete proposal came from a participant: a three-part
submission consisting of the code or other deliverable; AI interaction
logs, with the student directed to show how they contributed to the
result rather than simply dumping a transcript; and a reflection
document, submitted as video specifically to make it harder to generate.
Participants noted that making the process explicit in this way has a
secondary benefit---it goes some distance toward addressing inequities
in access to different AI tools, since the account of contribution is
what is being assessed rather than the polish of the output. The video
component did not go unchallenged: the facilitator noted afterward that
he had tried video deliverables himself and been troubled by students
using AI to script them---the same displacement the format was meant to
prevent, relocated one step upstream. The observation generalizes to
most of this section's proposals. Each moves the evidence somewhere the
model is not yet, and each invites the model to follow.

Convincing students to accept this was treated as a persuasion problem
with an honest framing. Weighting process means accepting some
short-term learning loss in exchange for later gains, and the argument
to students is that they will eventually hit a wall, and that the skills
need to exist before they reach it. Participants were clear that this
argument only lands from a particular footing: breaking down the
hierarchy barrier between instructors and students, positioning the
instructor on the student's side of the learning problem, and reframing
education as being for the student's development rather than a
compliance regime. Several described this explicitly as trying to break
the adversarial relationship that AI has intensified. A lighter note in
the same direction: AI also lets instructors build better and more
interesting assignments than they could before.

The facilitator's own Build Your Own Feature assignment was discussed as
a working example\footnote{\url{https://github.com/kevinlin1/huskymaps/blob/main/src/main/java/BYOF.md}}.
Its aim is not to assess fundamental learning objectives directly but to
integrate smaller learning objectives into a larger engineering design
lifecycle, so that process has somewhere to live structurally. Two
limitations were named by its own author. Fine-grained skills still
resist assessment this way, and may still require an invigilated,
controlled environment. And deploying the pattern across a curriculum
risks every course feeling the same, since the engineering design
lifecycle would repeat---acceptable while the field is experimenting,
less so as a permanent structure.

The room closed on what future assessment might actually look for. Not
whether the student produced the artifact, but whether they went through
a process, and how they coached the AI through it. Participants pointed
to the AI-Enhanced Developer profile in the appendix of Kam et
al.~(2025) as one attempt to characterize what that competence consists
of\footnote{\url{https://doi.org/10.1145/3696630.3727251}. Note that the
  appendix containing the AI-Enhanced Developer profile appears only in
  the authors' arXiv version, not in the version of record.}, and raised
structured decomposition as a candidate skill to assess in its own
right.

\pandocrule

\subsection{6. Rubric \& Grading
Redesign}\label{rubric-grading-redesign}

\textbf{Facilitator:} Shubbhi Taneja, Worcester Polytechnic Institute

\emph{This section covers both the Hour 1 strategy room and the Hour 2
question room ``What should rubrics and grading look like now?'', which
the facilitator summarized together.}

\subsubsection{The strategy}\label{the-strategy-5}

Changing what the grade is \emph{for}, rather than changing the
assignment. This covers rewriting rubrics to name AI use explicitly,
shifting weight between product and process, permitting AI subject to
disclosure, and treating AI literacy as a graded outcome in its own
right---making AI something students are assessed \emph{about}, not only
something they are assessed \emph{with}.

\subsubsection{Seeded questions}\label{seeded-questions-5}

\begin{enumerate}
\def\labelenumi{\arabic{enumi}.}
\tightlist
\item
  What actually changes when you add AI-awareness to a traditional
  rubric?
\item
  How do we grade or give credit for homework that involved AI, and what
  should it be worth relative to exams?
\item
  Can one rubric work for both students who embrace AI and students who
  avoid it?
\item
  How do we grade fairly when not every student has paid AI access?
\item
  How do we design rubrics that preserve the productive friction
  students need to learn, and don't just become a clean spec AI can
  fill?
\item
  Are our students, and are we, ready for process-based grading?
\item
  How does this translate to team projects, and to systems and
  networking courses?
\item
  Do we grade the product, the process, or both?
\end{enumerate}

\subsubsection{Discussion}\label{discussion-5}

Participants largely agreed that the balance is shifting toward
\emph{process}, though product still matters. Standards-based,
mastery-based and specs-based grading emerged as promising frameworks.
These approaches give students multiple opportunities to demonstrate
understanding and shift the emphasis from producing a single deliverable
toward demonstrating genuine learning. In specs-based grading, clearer
specifications can also improve the quality of the product itself.

The underlying challenge is that students tend to be outcome-oriented
and have pushed back on process-focused rubrics in some participants'
experiences. When AI can complete the deliverable, the deliverable alone
stops being meaningful evidence of learning.

Several participants therefore emphasized the need to establish
expectations early: What is the goal of the course? What counts as
cheating? When and why does process matter? One participant cited
Stephen Brookfield's \emph{Becoming a Critically Reflective Teacher},
recommending that instructors write a letter to students explaining
their teaching philosophy and then build a rubric that reflects it. This
raises a deeper concern, however: if students can use AI---particularly
agentic AI---to articulate or reconstruct their ``process,'' are
instructors actually assessing learning, or access to a polished AI tool
and the ability to prompt it effectively? This question makes the
distinction between visible process and genuine understanding especially
important, and it recurred in the fairness room as a matter of validity
rather than only fairness.

Participants offered a rich set of alternatives to traditional take-home
work, aimed at verifying that students actually understand what they
submit. Oral assessments were frequently suggested, including asking
students to explain code, walk through their reasoning, or identify and
diagnose bugs. Participants also noted the complications oral assessment
introduces for international and disabled students, and discussed
offering accommodations and language options directly to students who
seek them rather than advertising them to the class, so that the grade
reflects communication and process skills rather than English fluency.

Mastery checks and fluency checks, which need not be oral, were proposed
as lighter-weight alternatives where students demonstrate understanding
on demand. Other formats included whiteboard exams, in-class debugging
exercises and multi-stage group projects where students review and fix
each other's work; one participant reported a debate over programming
languages working particularly well. A broader principle emerged:
high-stakes credit should rely primarily on work that can be directly
verified, while take-home assignments can function as practice for those
assessments.

When AI use is assumed or permitted, participants suggested redesigning
assignments and rubrics to evaluate the thinking behind the product.
Possible requirements included citing AI sources, submitting prompts,
comparing outputs from different AI tools, and documenting decisions
through reflections, analytical questions, design reasoning, or non-code
artifacts such as diagrams. Participants also cautioned against assuming
that AI-generated artifacts are necessarily better---one reported that
database diagrams generated by AI were less accurate than what students
had produced before. Rubrics should reward understanding and substantive
decision-making rather than polish alone. More demanding
specifications---such as requiring students to measure code throughput
or performance---can also push assignments beyond what AI can easily
generate. One participant raised the idea of ``hidden rubrics'':
undisclosed evaluation criteria that make it harder for students to
optimize their work directly against the rubric.

Another structural approach was tiered grading. A baseline rubric could
bring everyone to a defined grade ceiling, while optional additional
assessments---including potentially AI-integrated work---could allow
students to reach higher grades. This preserves student choice without
penalizing those who choose not to use AI, a group participants took
seriously: reasons given for refusal included environmental impact,
objection to having one's data used for training, and other ethical
concerns. Participants also noted that instructors do not always have
institutional freedom on this question, and that students receive mixed
messages from different faculty. At the same time, if AI literacy is
genuinely part of the discipline, participants argued that it should be
named explicitly as a learning goal in the syllabus. Framing AI as
something students learn about, rather than merely something they learn
with, makes its inclusion more intentional and pedagogically grounded.

Finally, participants discussed AI as a tool for handling lower-level
tasks so that class time and graded work can focus more heavily on
analysis, evaluation, design, and other higher-order thinking. That
shift, however, requires assessments that deliberately target those
skills rather than rewarding completion of tasks AI can perform easily.

Two questions remain unresolved. First, if AI can increasingly generate
not only products but convincing accounts of the process behind them,
what forms of assessment provide reliable evidence of genuine learning?
Second, if students are expected to use AI to move more quickly to
higher-order work, how much lower-level practice and fluency do they
need first, in the light of Bloom's taxonomy? The challenge is not
simply deciding whether AI belongs in assessment, but determining what
students must still do themselves for learning to occur---and how
instructors can reliably tell that it has.

\pandocrule

\subsection{7. Oral \& Interactive
Assessment}\label{oral-interactive-assessment}

\textbf{Facilitator:} Ranysha Ware, Swarthmore College

\emph{This section covers both the Hour 1 strategy room and the Hour 2
question room ``How do we assess richly at scale?''}

\subsubsection{The strategy}\label{the-strategy-6}

Assessment conducted through live interaction: oral exams,
interview-style checkpoints, code walkthroughs, presentations, and
demos. The student is present and answering, which makes the assessment
robust to how the submitted artifact was produced. These rooms drew the
most detailed practitioner experience of any in the workshop, and also
the most consistent single objection: scale.

\subsubsection{Seeded questions}\label{seeded-questions-6}

Posed and worked through in the Hour 1 room, in this order:

\begin{enumerate}
\def\labelenumi{\arabic{enumi}.}
\tightlist
\item
  What does a simple, robust rubric for an oral exam or online code
  walkthrough look like, one that targets the learning we actually want?
\item
  How do we make oral assessment scale to classes of 30, 100, or more?
\item
  Can peer or UTA assessors help carry the load, and how do we keep
  grading consistent, especially without TA or grader support?
\item
  How do we probe real comprehension when a student is presenting
  AI-generated code?
\end{enumerate}

Seeded but never posed, because the room worked through the list in
order and ran out of time:

\begin{enumerate}
\def\labelenumi{\arabic{enumi}.}
\setcounter{enumi}{4}
\tightlist
\item
  How do we tell an honest low grade from an academic-integrity case,
  and what artifacts do we keep if a grade is challenged?
\item
  How do we run oral exams that aren't just whiteboard interviews, for
  example having students modify their own homework in real time?
\item
  Where can AI help run or support oral assessments, and what breaks?
\end{enumerate}

From the Hour 2 room:

\begin{enumerate}
\def\labelenumi{\arabic{enumi}.}
\tightlist
\item
  How do oral, interactive, or authentic assessments scale to hundreds
  of students, and what do we give up at scale?
\end{enumerate}

Questions 5 through 7 are listed because they are part of the record of
what this room's registrants wanted to discuss. Substantial material
bearing on all three appears below, but it arrived through the earlier
questions rather than because the room took them up directly.

\subsubsection{Discussion}\label{discussion-6}

This room was run tightly to its question list: the facilitator posed
each seeded question in turn, let discussion run, then moved on. That
structure is why the material below can be traced to particular
questions, and also why the integrity discussion appears where it
does---it surfaced under the comprehension question rather than under
the integrity question, which was never asked.

On rubrics, the room's experience converged on scoring the \emph{quality
of the account} rather than the artifact. One participant running
monthly oral assessments with twenty students prepares a script in
advance with a rubric attached to each question, and does not grade code
directly at all. A CS2 course assesses students in groups of three on a
five-point scale where five means working through the problem with no
prompting from the proctor and one means getting through it with
substantial help from teammates---the dimensions being how well a
student can explain the problem without help, and how precisely. Where
code is involved, questions are generated from the student's own
implementation. Two participants shared working rubrics into the
room\footnote{Both were shared into the room by participants, and are
  linked here with their authors' permission:
  \url{https://cmu.box.com/s/p7nd2bju11f35pcaub8qg0q284b1giej} and
  \url{https://drive.google.com/file/d/1ghhnsgdBIHhfpiFT5ZaQhGoePmtS1GuS/view}.}.

Question reuse surfaced immediately: any question set leaks once the
first cohort sits it. One participant's response was to share questions
ahead of time, a practice they had already used for a writing-based
assessment, on the reasoning that a memorized answer is detectable in
conversation. The same participant described the format as robust but
brutal. Students who had done their own work answered within thirty
seconds and knew exactly where to navigate in their code; students who
had not done their own work spent the full ten minutes on the first
prompt, searching through their own submission. Watching that happen,
and resisting the urge to help, was reported as genuinely painful.

Scale came up in both rooms, though how sharply varied with course size:
for some participants it was the governing constraint, for others a
secondary concern. When it was discussed, the approaches participants
had actually run were all forms of distributing or sampling the work.
Peer assessment, with groups presenting to other students and the
instructor steering conversations that drift toward leniency, capped at
forty percent of an assignment's weight against a final exam carrying
sixty. A pre-GenAI software engineering course ran in-person code
reviews for two hundred students, interviewing each team member about
their individual contributions; scaling it meant distributing reviews
across ten to fifteen TAs who first sat in together on the opening round
and sample-reviewed prior years' projects to calibrate, then handled
three or four teams each with a staff debrief afterward. Sampling was
the other lever: one participant keeps small-group work but makes the
deliverable an in-person oral assessment taken without the team, calling
only a subset of students in any given week so no one knows in advance
whether it is their turn. Another allocates twenty minutes per student
for a single question---design an algorithm and prove its
correctness---and reports that twenty minutes is enough to tell, even
when a student needs longer to finish explaining.

Consistency without a calibrated TA pool was the harder version of the
problem, and participants accepted inconsistency on a semester-end
project more readily than on weekly work. For written work the practice
described was a standing calibration meeting where the grading team
marks independently then convenes to reconcile. Translating that to oral
assessment is harder: the suggestions were recording sessions with
consent and reusing them across semesters, withholding all grades until
every student has sat the assessment, and running role-play sessions in
staff meetings---which participants noted has value even where it fails
as calibration, because it gets TAs comfortable conducting an oral
assessment at all.

The room was most interesting on what an oral assessment is actually
\emph{for} when the code was AI-generated. One participant described a
student who had clearly had AI write the code but could answer every
question about it and evidently understood it, and felt the grade should
be lower without being able to say why. The room's response was that
this depends entirely on permission: if AI use was allowed and the
student used it well, that deserves credit rather than suspicion; if it
was not allowed, the grade follows from the policy violation, not from
the oral performance. Separating a poor performance from misconduct was
treated as a distinct problem, and consequential, since misconduct means
filing with an office. Participants reported that the distinction is
usually clear when the oral assessment is read \emph{against} the
artifact: a student submitting incomplete code who does not realize they
don't understand it looks nothing like a student submitting flawless
code who cannot explain what a variable is. One participant recounted
asking a student why their code contained emoji characters---which
keyboard key produced that?---and getting no answer.

Two design principles followed. An oral assessment should not consist
only of explaining a submitted artifact, because that can be rehearsed;
it should require extending the code or making a change in real time, or
at minimum asking what the student would do to accommodate a new
requirement. And where misconduct cannot be established, the workable
response is structural rather than punitive: downweight the artifact,
increase the weight on the oral assessment. Participants also noted a
broader disposition---a student who put in the effort to genuinely
understand what AI produced has learned something, whatever the ethics
of how they got there, and policing every case is not the job.

The scale room extended this into institutional and legal constraints
participants had hit rather than theorized. Distance and short courses
of fifty to seventy students make synchronous oral assessment across
time zones impractical. Testing centers change the question from whether
assessment scales to what belongs in it: one participant runs four
fifty-minute exams with students choosing their own sitting times, and
reported a five-point drop in scores between semesters alongside
complaints about not finishing in the same time---their hypothesis being
that the gap reflects assignments now being done with AI. Remote
proctoring has shifted from a human watcher plus lockdown browser to
recorded sessions with AI flagging and human review, which participants
observed has increased rather than decreased the review burden---and
recording is unavailable in some jurisdictions, where GDPR prohibits
capturing biometric data. Global cohorts push toward two sittings split
by hemisphere, or delegating final exams to physical centers in each
country under national rather than international law. Who may grade
varies too: in the UK, TAs cannot.

On AI's own role in assessment the room split. Some described local
models generating per-student assessments, and question generators that
analyze a student's artifact and produce questions from it with TAs
marking the answers---not implemented, on grounds of time and risk.
Another participant declined the direction on principle, holding that
the human value added in grading is the point. The closing observations
were diagnostic rather than prescriptive: the same student scoring in
the seventieth percentile on an oral assessment and the ninetieth on
homework; the sense that a struggling student registers as the
instructor's failure; and the framing that we assess answers when what
we want to assess is learning, which we can only reach through students'
responses.

\pandocrule

\section{Part II: Cross-cutting Question
Rooms}\label{part-ii-cross-cutting-question-rooms}

\subsection{8. Fairness \& Trust}\label{fairness-trust}

\textbf{Facilitator:} Fraida Fund, NYU Tandon School of Engineering

\subsubsection{The question}\label{the-question}

\emph{How do we keep assessment fair and rebuild trust?} This room was
designed as cross-cutting rather than tied to a single strategy, on the
observation that questions about integrity, unequal access, and the
instructor-student relationship surfaced in registrants' submissions
across every strategy.

\subsubsection{Seeded questions}\label{seeded-questions-7}

\begin{enumerate}
\def\labelenumi{\arabic{enumi}.}
\tightlist
\item
  How do we distinguish an honest performance issue from an academic
  integrity case?
\item
  How do we handle unequal access to AI tools?
\item
  How do we rebuild trust without turning assessment into surveillance?
\end{enumerate}

\subsubsection{Discussion}\label{discussion-7}

The discussion focused on how to assess student learning fairly when AI
can generate much of what instructors previously evaluated directly.
Participants questioned whether traditional learning outcomes remain
appropriate and whether grades still measure what educators intend them
to measure. As technical work changes, qualities such as problem
decomposition, iteration, testing, judgment, and responsibility for
results may become more important than code correctness alone. However,
instructors still need assessment methods that are reliable, defensible,
and capable of distinguishing genuine learning from work that has been
substantially offloaded to AI.

Several participants have revised assignments and rubrics to place more
weight on process. Students may submit AI interaction logs, describe how
code evolved, document testing and edge cases, or reflect on how they
used AI---one participant described a rubric splitting credit roughly in
thirds. These records can encourage responsibility and reveal elements
of students' problem-solving processes, but participants generally
viewed them as metacognitive and informational tools rather than
reliable evidence for grading. Students may omit AI use, use paid tools
unavailable to their peers, interpret ``AI use'' inconsistently, or
report only low-risk uses. One participant observed that students no
longer recognize some things as AI at all---search results with
generated answers embedded, for instance---and another that students
default to declaring no AI use simply to avoid writing the details.
Finding a way to ask that does not sound antagonistic was an open
problem for the room.

The discussion also raised concerns about what students may no longer be
learning. Some instructors observed strong project outputs alongside
declining performance on written assessments of foundational material.
This creates an ongoing need to revisit learning objectives, especially
because AI tools and industry expectations are changing faster than
curricula can be revised---and, as one participant noted, industry does
not know what it wants either. Grades nevertheless remain important
signals for graduate programs and employers, increasing the pressure to
ensure that they represent something credible and comparable.

In response, some instructors are shifting greater weight toward
controlled assessments, including handwritten exams, in-class quizzes,
computer-based testing facilities, oral examinations, demonstrations,
and individual meetings. These formats can provide stronger evidence of
individual understanding, but they consume time, are difficult to scale,
and may increase anxiety. Open-book exams, retakes, and
application-focused questions were discussed as ways to reduce stress
without sacrificing integrity; one participant described allowing
unlimited paper notes, teaching note-making as a study strategy, and
designing questions that cannot be answered directly from notes.
Controlled testing environments may also be less antagonistic than
ordinary classroom proctoring: they can reduce opportunities for visible
cheating while making students less concerned that innocent behaviors
will be interpreted as suspicious.

At the same time, many important outcomes---such as extended design,
collaboration, and project work---cannot be assessed adequately in a
conventional exam, and proposals to treat computing more like a
laboratory science introduce their own questions about supervision and
isolation. One participant also noted a reluctance to give up class time
for quizzes at all.

Underlying these challenges was a concern about rebuilding trust. Most
students are believed to want to learn, but a small number who use AI in
ways that counter the goals of the course can undermine instructors'
confidence and create resentment among students who follow the rules.
Honest students may lose trust when they see cheating go unaddressed,
while increased surveillance, the elimination of take-home assessments,
and repeated warnings about misconduct can make all students feel
distrusted. One participant described being perceived as the instructor
who does not trust students precisely because she had moved work into
class and one-on-one meetings; another described announcing before an
exam that cheating would be handled afterward rather than confronted in
the room, intended to reassure honest students but with the unintended
side effect of signalling that cheating was expected.

Participants emphasized clearly communicating permitted and prohibited
uses, beginning from an assumption of trust, responding transparently
when that trust is broken, and holding students accountable without
turning assessment into an adversarial process. The central tension is
how to acknowledge the lower barrier to misconduct and the high stakes
surrounding grades while preserving a collaborative relationship in
which students can be honest about their AI use and instructors can
remain confident in what grades represent. Participants situated part of
this beyond the classroom: the cost of education and the competitiveness
of what comes next have created a high-stakes environment that neither
students nor instructors chose.

\pandocrule

\subsection{9. Motivation}\label{motivation}

\textbf{Facilitator:} James McGuffee, Columbia College (Missouri)

\subsubsection{The question}\label{the-question-1}

\emph{How do we keep students motivated to learn?} Like Fairness \&
Trust, this room was designed as cross-cutting. It was the only room to
take student motivation as its primary object rather than as a
constraint on assessment design.

\subsubsection{Seeded questions}\label{seeded-questions-8}

\begin{enumerate}
\def\labelenumi{\arabic{enumi}.}
\tightlist
\item
  How do we motivate real effort when AI can produce the answer?
\item
  How do we align process or authentic assessment with intrinsic
  motivation?
\item
  What makes students want to learn the hard parts anyway?
\end{enumerate}

\subsubsection{Discussion}\label{discussion-8}

The room began by turning the question backward: how did we know
students were motivated \emph{before} AI? The answers participants gave
for their own students were varied enough that several concluded the
question resists a single answer. Employment and career prospects came
up first and most often, along with a concrete tactic---giving students
exam questions that mirror job interview questions, so the connection
between coursework and hiring is visible rather than asserted. Others
reported students citing gaming, or problem-solving for its own sake.
Asked what had motivated \emph{them}, participants named the fun and joy
of learning, with the implication that this is not what most of their
students are working from.

To get a rough read on the mix, one participant put a four-way breakdown
into the chat: students who are fully self-motivated and love learning
computing regardless of grades; students who are interested but need
grades and structured assignments to stay on track; students who are
barely or purely extrinsically motivated and respond only to penalties
or job prospects; and students who are unmotivated and will cheat or
avoid work at any opportunity. Two participants offered estimates:
10/20/50/20 and 10/60/20/10. The estimates disagree substantially about
the middle of the distribution but agree that the fully self-motivated
group is small---roughly one student in ten---and that the great
majority depend on some combination of career prospects, structure, and
external nudges.

That framing led to the room's central practical thread: ownership.
Participants argued that students need more ownership of their learning,
and acknowledged that the profession did not do a good job of explaining
\emph{why} it teaches what it teaches even before AI made the question
urgent. Asked how ownership is actually built, the suggestions were
small and personal rather than structural: explicitly explaining the
purpose of what is being taught, and creating room for reflection
through brief private conversations with individual students. One
in-person strategy discussed was breaking students into small groups for
meta conversations about the value of learning itself, rather than
addressing the topic only from the front of the room.

Participants also questioned whether motivating students is faculty's
job at all. The case against is that students carry many courses, that
grades are a necessary but insufficient motivator, and that
responsibility has to sit somewhere with the student. The case for
emerged from a broader worry the room voiced: if a student can simply
have a conversation with a chatbot, what is the university for?
Participants' answer was that students learn by interacting with each
other, and that this holds even for high-achieving students---which
locates part of the value of the institution in something a model cannot
supply, and makes fostering that interaction a legitimate faculty
responsibility. A related fear was named directly: that grades are
becoming the only motivational tool anyone has left.

The room spent its remaining time on the student question that motivated
the session---\emph{if I'm going to use AI in the future, why do I need
to know how to do this myself?}---and rejected the premise rather than
the question. Several participants argued that problem-solving remains a
valued human skill regardless of who writes the code. The analogy that
recurred was the calculator: arithmetic did not stop being taught, and
did not stop being necessary, because a machine could do it. The deeper
objection participants raised was epistemic. Students don't know what
they don't know when using AI, and a model can misrepresent the state of
a student's technical understanding back to them---producing confidence
that has no underlying competence to support it. Students are also,
participants noted, poorly positioned to see the long-run benefits of
understanding, which makes the argument for foundational work hard to
win on the students' own terms. Related to this, participants raised
concern about students' reading ability more generally, observing that
students need to be able to read and understand their AI conversations,
not merely produce and accept them.

The room's conclusion was that the productive move is to challenge the
assumption that AI can do everything. Framed as a tool that produces
both good and bad results, AI stops being a reason foundational learning
is obsolete and becomes a reason it is necessary: students need enough
technical fluency to catch errors, debug real systems, and evaluate what
they are handed. Helping students recognize their own blind spots was,
in the room's account, the precondition for everything else.

\begin{quote}
\textbf{A note on this room's record.} Partway through the session a
participant put the room's three seeded questions to a language model
and pasted its approximately 900-word reply into the shared document.
The reply is not summarized above, and the summary that had been drafted
from those notes has been set aside; this section is written from the
human discussion only. We record the episode because a room on
motivation reaching for a model mid-conversation is itself a datum about
the moment this workshop took place in.
\end{quote}

\pandocrule

\section{Part III: Themes Across the
Workshop}\label{part-iii-themes-across-the-workshop}

Some arguments surfaced in rooms that had no reason to be discussing the
same thing, and a few patterns are visible only from above the
individual sections. Both are collected here rather than repeated
section by section.

\subsubsection{Do students still need the fundamentals, and what are
they
now?}\label{do-students-still-need-the-fundamentals-and-what-are-they-now}

This was the most widely distributed disagreement in the workshop. It
arrived independently in the open-ended task design rooms, the
controlled assessment rooms, the critique room, and the motivation room,
and it was not resolved in any of them.

The case for continuity was put most sharply in the open-ended design
room. A participant observed that expert programmers who claim they no
longer write code always, eventually, describe a moment where they
inject a highly specific technical judgment---use this library rather
than that one---and that judgment is unavailable without deeply
understanding what those libraries do. On that reading students still
need most of the traditional curriculum: how memory works, what
execution actually means, why assembly is taught at all. The same
intuition appeared elsewhere in different clothing. The critique room
asked whether a student can meaningfully critique code without being
able to write it, and answered mostly no, reaching for the analogy of a
music critic who needs theory. The motivation room reached for the
calculator: arithmetic did not stop being taught because a machine could
do it. And a participant in the critique room gave the argument its
professional form---engineers are expected to evaluate, debug, test, and
justify solutions whatever their origin, and none of that is possible
without understanding the fundamentals of coding itself.

The case for change came in two quite different versions, which are
worth separating because they are often conflated. The first is
empirical: current models already do the thing the continuity argument
says they cannot. They will pull both libraries, benchmark them, write
the tests, and decide, in minutes and at a precision humans struggle to
match. The second is not about capability at all. Writing code by hand
is mentally taxing in a distinctly unpleasant way, and arguably
dehumanizing---it forces a person to address a machine in the machine's
language rather than a human's. If that barrier falls away, more people
may be able to do computational work, and the space it vacates could go
to things the curriculum has never had room for. Most computing
students, one facilitator observed, have never built anything they would
actually use; the canonical exercise is reversing a linked list.

Where rooms tried to name what specifically survives, they produced
overlapping lists rather than answers. The controlled assessment room
offered foundational concepts, computational thinking and problem
solving, AI literacy and prompting, competence with platforms, and code
comprehension---with less emphasis on building from scratch and more on
verifying and adapting, and a tentative suggestion that memorization may
no longer belong at all. The open-ended design room offered loop
invariants and algorithmic correctness, which computing has never really
taught and which would return the curriculum to Dijkstra's insistence
that students be trained as mathematicians; program structure and
design, with one participant proposing a deliberate throwback to
flowcharts on the grounds that not writing the code makes understanding
its shape more important rather than less; ``taste,'' borrowed from a
recent Titus Winters talk and immediately complicated by the observation
that taste is unreachable without knowing what good looks like; and
systems knowledge, since graduates will manage codebases where every
change has consequences for memory, bandwidth, and behavior across
environments. The ambitious projects room added a different register:
reading and evaluating unfamiliar code, articulating objectives,
decomposing work, giving useful feedback, and managing an agent's
progress---skills closer to supervising a collaborator than to
programming.

Two observations cut across the disagreement. The first is an asymmetry
named explicitly in the ambitious projects room: educators built their
foundations before these tools existed and learned to work with them
afterward, a sequence today's students cannot reproduce, which leaves
educators with no firsthand basis for judging which struggles remain
necessary. The second is that AI's benefit appears to accrue fastest to
students who need it least. The controlled assessment room concluded
that AI helps most those who already understand the material or have
solid computational thinking, and that generative AI produces excellent
personalized tutorials which weaker students need training before they
can use. The motivation room reached the same place from the other
direction: students do not know what they do not know when using AI, and
a model can misrepresent the state of a student's understanding back to
them, producing confidence with no competence underneath.

\subsubsection{Does code remain something humans work
with?}\label{does-code-remain-something-humans-work-with}

This thread deserves separate treatment because it kept escalating.
Participants would arrive at a position, and then someone would point
out that the position assumed something that was itself in question.

It began as a claim about emphasis: students will read far more code
than they write. That premise was widely shared, and the critique room
built a curricular argument on it, proposing that reading and critiquing
code be introduced early and developed progressively as the skill that
survives when writing does not.

The open-ended design room then took the premise apart. One participant
described designing a Reading Code special-topics course the previous
spring on exactly that basis, then said that six months later she no
longer believed the skill would be needed: she had just spent a week
having one model write code, a second review the pull request, and the
first apply the fixes, with no human reading anything. If code review is
passing between models, reading is not obviously the durable skill
either.

From there the question went one level deeper, to whether code remains a
human-facing artifact at all. A facilitator invoked the moment in the
AlphaGo documentary where a system stops learning from human play and
starts learning from itself, and suggested models may eventually
maintain code in forms optimized for models rather than
people---describing having told a model he would never read a particular
codebase and it should organize the work for its own use. Pushed
further, the speculation was that models may eventually emit artifacts
no human is expected to read at all, decompiling them on demand into
whatever language a human happens to want when something goes wrong. The
critique room arrived independently at the same question from a
different direction, closing its discussion by asking whether LLMs will
simply develop their own codebase, easier for them to produce and for
non-developers to understand.

If that is where things land, the object of computing education shifts
from an artifact to a relationship. The ambitious projects room had
already described the required competence in those terms: articulating
objectives, decomposing work, evaluating results, giving useful
feedback, and managing an agent's progress---closer to supervising a
collaborator or a research team than to programming. On that account the
question is not which code skills survive but whether code is the right
unit of instruction.

Three counterarguments were offered, and they are not equivalent.
Resilience: someone has to understand the systems if the tools become
unavailable, the same reason we still have hammers and repair
technicians. Economics: LLM use is currently subsidized, and when it is
not, understanding code will separate efficient use from burning tokens
to no purpose. Both concede the direction of travel and argue for
retaining capability against contingencies. The third does not. Code,
one participant argued, is a language deliberately designed to capture
semantics with a precision natural language lacks---the reason we write
programs rather than describe them was never that machines could not
understand English, but that English is ambiguous. Models have become
good at resolving that ambiguity statistically, but resolving it is not
the same as eliminating it, and a person who cannot read the precise
form has no way to check what was inferred from the imprecise one. That
is an argument that code remains necessary, not merely prudent.

The thread is unresolved, and it bears directly on the rest of this
report. Critique of AI output, code walkthroughs, and oral assessment
built around explaining or extending a submitted artifact all presuppose
that working with code is a skill worth verifying. If that
presupposition weakens, a substantial portion of the assessment
strategies gathered here are transitional.

\subsubsection{Scale is the binding
constraint}\label{scale-is-the-binding-constraint}

Almost every remedy proposed in this workshop works at small scale and
degrades at large scale, and participants knew it. The pattern was so
consistent that it is better understood as a structural feature of the
current moment than as a complaint.

Oral assessment gives the strongest evidence of individual understanding
and is the least scalable thing anyone proposed. Critique of AI output
was raised alongside the figure of 400-plus students and a handful of
TAs. Process evidence requires someone to read it. Individual meetings,
demonstrations, and code walkthroughs all consume instructor time in
direct proportion to enrollment. Even controlled assessment, the most
scalable option discussed, ran into the problem that paper-based exams
are painful to grade and that large AI-free assignments will not fit
inside class time.

The responses that participants had actually implemented were all forms
of distributing or sampling rather than solving: peer assessment capped
at a fraction of the grade, TA pools calibrated together before a first
round of reviews, sampling a subset of students each week so that nobody
knows when their turn comes, testing centers that decouple assessment
from class time, automated grading of code artifacts with human effort
reserved for smaller pieces. Two participants proposed handing
assessment itself to AI---local models generating per-student questions
from their submitted artifacts, with TAs marking the answers---and
another declined the idea on principle, holding that the human judgment
in grading is the point.

The uncomfortable implication is that the strategies best supported by
evidence of learning are the ones least available to the institutions
teaching the most students. Several participants named this directly as
an equity problem between institutions rather than between students.

\subsubsection{Process evidence is valued but not
trusted}\label{process-evidence-is-valued-but-not-trusted}

Across three rooms, participants converged on the same position from
different starting points, and it is sharper than the enthusiasm for
process-based assessment might suggest.

Instructors are collecting process evidence: AI interaction logs,
accounts of how code evolved, testing and edge-case documentation,
reflections, disclosure statements. When asked directly what they use it
for, participants in the fairness room were unanimous that they do not
grade it. It functions as a metacognitive prompt for the student, a
signal that AI use is permitted provided the student owns the output,
and a source of insight for revising the course next time.

The reasons for not grading it are practical and hard to design around.
Students omit uses, use tools their peers cannot access, interpret ``AI
use'' inconsistently, or disclose only the low-risk uses. Some no longer
recognize certain things as AI at all. Some default to declaring no use
simply to avoid writing the details. And the deeper objection, raised in
the rubrics room, is that agentic AI can generate a plausible account of
a process that never happened---so a process-focused rubric may be
assessing access to a good tool and the ability to prompt it.

The most concrete proposal in the workshop, a three-part submission of
deliverable, interaction logs, and a video reflection, drew exactly this
objection from the facilitator who reported it: he had tried video
deliverables himself and found students using AI to script them. The
displacement moves upstream rather than stopping. Nobody in the workshop
identified process evidence that resists being generated, which leaves
an uncomfortable gap between how much participants valued process and
how little they were willing to stake a grade on it.

\subsubsection{Unequal access is a validity problem, not only a fairness
one}\label{unequal-access-is-a-validity-problem-not-only-a-fairness-one}

Unequal access to paid AI tools was raised in at least four rooms, and
it arrived twice in each: first as a matter of fairness, then as a
matter of what a grade means.

The fairness version is familiar. Students without paid subscriptions
are disadvantaged on assignments where capability scales with the tool,
and participants discussed institutional model access, ``bring your own
provider'' agent harnesses, free-tier and self-hosted models, and
NSF-supported research infrastructure as mitigations. A related case was
raised sympathetically in two rooms: students who decline to use AI on
ethical, environmental, or privacy grounds, for whom the mitigation is a
less capable model and therefore a worse result.

The validity version is more serious and less often stated. If
assessment weights process evidence, and process evidence is easier to
produce convincingly with a better tool, then the assessment is partly
measuring purchasing power. One participant put it as not being able to
distinguish students who used AI for trivial things and said so from
students who used it more and did not. Another noted being unable to
tell whether students had used tools beyond those the institution
provided. In both cases the problem is not that some students are
disadvantaged but that the instructor cannot say what the resulting
grade represents.

\subsubsection{Difficulty is now a moving
target}\label{difficulty-is-now-a-moving-target}

Participants repeatedly described calibrating assignments against model
capability and finding the calibration expire. A cybersecurity
instructor described problems that required genuine problem-solving with
AI assistance seven or eight months earlier and are now simply solved by
it, having already raised the difficulty once. A systems instructor who
ran an open-ended project observed that models had since become better
at generating tests and handling a whole package, and expected an agent
could now complete her specification end to end. One room noted that
models were not good at producing even small programs two years ago. An
assignment built around a specific model limitation---one instructor
exploited the observation that models handled two coupled concerns
poorly---is calibrated to a version, not to a task.

The consequences run in two directions. Anything built to resist AI has
a shelf life measured in semesters, which one facilitator estimated at
roughly one. And the difficulty of an assessment is now partly a
function of the quality of prompt a student can write, which makes
``hard for AI'' a property of the interaction rather than of the
question.

This is what makes the curricular-response argument in the open-ended
task design section hard to act on. Curricular revision operates on a
timescale of years, and the thing being revised against is moving faster
than that.

\subsubsection{Almost everything here was done by someone acting
alone}\label{almost-everything-here-was-done-by-someone-acting-alone}

A pattern is visible across the report that no single room could have
seen. Nearly every change documented in these pages was made by an
individual instructor, in a single course, without institutional
coordination and often against local friction.

Participants described colleagues within the same department taking
opposite positions---some permitting AI freely as another tool, others
requiring students to separate their own contributions from the
model's---with no departmental position reconciling them, and students
receiving the resulting mixed messages. Some reported not having
institutional freedom to decide at all. One noted that their institution
does not permit disciplinary measures for AI use; another that theirs
permits them only for unauthorized use, and that realism is still
required. An instructor who moved assessment into class and one-on-one
meetings reported being perceived as the one who does not trust
students, precisely for taking the problem seriously.

The argument for a curricular rather than a course-level response was
made explicitly in the open-ended design room and went essentially
unopposed. What was also unopposed was the observation that curricular
conversations, where they have started, are not expected to conclude
quickly enough to help students currently enrolled. So instructors
continue making local decisions and absorbing local costs.

That may be the most consequential finding in this report. The
strategies catalogued here are not the output of a discipline deciding
how to respond. They are the output of individuals improvising, in
parallel, mostly without knowing what their colleagues are doing---which
is, in the end, the reason a workshop like this one had something to
gather.

\subsection{Limitations}\label{limitations}

This is a record of a conversation, and it should be read as one.

Participants were self-selected twice over: they chose to register for a
workshop about adapting assessment for AI, and then chose which room to
enter. The result over-represents educators already convinced that
something must change, and under-represents those who have concluded
that little needs to. It skews toward people with the institutional
latitude to redesign a course. Where instructors described constraints
they could not overcome, those constraints are reported here; where
instructors were unable to attend at all because of them, they are
invisible.

Nothing here is measured. Reports of what worked are self-reports,
mostly unaccompanied by outcome data, and often about a single offering
of a single course. The one participant who cited a numeric result---a
five-point drop in exam scores between semesters---offered a hypothesis
about the cause rather than a finding.

Room composition was uneven. Some rooms had twenty participants and
skipped spoken introductions; others had four and worked through every
question. The sections in this report are of comparable length by
editorial decision, not because the rooms were of comparable size or
depth.

Finally, this report is a snapshot of late July 2026, and several of its
central disagreements turn on what models can currently do. Participants
themselves repeatedly observed that positions taken six months earlier
had not survived. It would be surprising if all the positions recorded
here survive six months either.

\subsection{About This Report}\label{about-this-report}

\textbf{Method.} Each room had a tab in a shared, participant-editable
document in which its facilitator took notes. The session was not
recorded; Zoom's live transcript ran to support note-taking, and some
facilitators kept fuller transcripts of their own rooms. Sections were
drafted from those notes and transcripts by the working group, then
reviewed by the facilitator of each room.

\textbf{Anonymity.} The record is anonymous by default. No idea in this
report is attributed to a named individual. Institutional details,
course identifiers, and details of individual student cases have been
removed or generalized where they would identify a person. Two teaching
resources shared into a room are linked with their authors' permission.

\textbf{Authors and contributors.} This report has two lists, both
alphabetical by surname. The nine working group members appear as
authors: they convened the workshop, facilitated its rooms, and wrote
this report. Workshop participants who wish to be acknowledged appear as
contributors on the title page.

Contributor listing is opt-in and applies to the report as a whole.
There is no idea-level attribution, nothing in this report is traceable
to a named participant, and declining has no effect on anything else.
The list was collected after this draft circulated to attendees. It
reflects opt-ins received to date and stays open. Participants may still
request to be acknowledged.

\textbf{Continuing the conversation.} Attendees were invited to continue
this discussion at \url{https://computingeducators.org}.

\textbf{Citation.} Akbar, M. S., Challen, G., Fund, F., Hopkins, C.,
Karnalim, O., Lin, K., McGuffee, J., Taneja, S., and Ware, R. (2026).
\emph{AI Can Do Your Homework. Now What? Report from an online workshop
on computing assessment in the age of generative AI.} SIGCSE Virtual
Working Group. \emph{{[}identifier pending{]}}

Authors are listed alphabetically by surname; all nine facilitated or
supported rooms at the workshop and contributed to this report.
Contributors are acknowledged separately, also alphabetically, and are
not authors.

\end{document}